\documentclass[a4paper,reqno,12pt]{amsart}

\usepackage{amsmath,amsthm,amssymb,color,graphicx}

\theoremstyle{plain}

\theoremstyle{definition}

\theoremstyle{remark}

\numberwithin{equation}{section}
\numberwithin{theorem}{section}
\numberwithin{figure}{section}
\numberwithin{table}{section}

\DeclareMathOperator{\lcm}{lcm}

\newcommand{\cA}{{\mathcal A}}

\newcommand{\cC}{{\mathcal C}}

\newcommand{\cF}{{\mathcal F}}

\newcommand{\cI}{{\mathcal I}}

\newcommand{\CC}{{\mathbb C}}
\newcommand{\NN}{{\mathbb N}}

\newcommand{\ZZ}{{\mathbb Z}}

\newcommand{\dCo}{\text{h}^\vee}
\newcommand{\Co}{\text{h}}

\newcommand{\func}[2]{#1 \left( #2 \right)}
\newcommand{\set}[1]{\left\{ #1 \right\}}
\newcommand{\bilin}[2]{\left( #1  |  #2 \right)}

\newcommand{\pardiff}[2]{\frac{\partial #1}{\partial #2}}

\newcommand{\fg}{{\mathfrak g}}
\newcommand{\wfg}{\widehat{\fg}}
\newcommand{\wLa}{\widehat{\Lambda}}

\begin{document}

\title[D-branes on group manifolds and fusion rings]
      {D-branes on group manifolds and fusion rings}

\author[P Bouwknegt]{Peter Bouwknegt}

\address[Peter Bouwknegt]{Department of Physics and 
Mathematical Physics, and Department of Pure Mathematics \\
University of Adelaide \\
Adelaide, SA 5005 \\
Australia}
\email{pbouwkne@physics.adelaide.edu.au, 
       pbouwkne@maths.adelaide.edu.au}

\author[P Dawson]{Peter Dawson}

\address[Peter Dawson]{Department of Physics and Mathematical Physics\\
University of Adelaide \\
Adelaide, SA 5005 \\
Australia}
\email{pdawson@physics.adelaide.edu.au}

\author[D Ridout]{David Ridout}

\address[David Ridout]{Department of Physics and Mathematical Physics\\
University of Adelaide \\
Adelaide, SA 5005 \\
Australia}
\email{dridout@physics.adelaide.edu.au}


\thanks{PB was supported by a Senior Research 
Fellowship from the Australian Research Council, and would also like to 
acknowledge financial support and hospitality of the CITUSC Center
for Theoretical Physics during various stages of this project.\\
ADP-02-92/M101, [{\tt arXiv:hep-th/0210302}]}

\begin{abstract}
In this paper we compute the charge group for symmetry preserving 
D-branes on group manifolds for all simple, simply-connected,
connected compact Lie groups $G$.
\end{abstract}

\maketitle

\section{Introduction} \label{secintro}

D-branes on group manifolds have been the subject of intensive 
research over the past few years (see, e.g., 
\cite{KS}--\cite{IS}), as they
provide a good laboratory for testing our intuitive understanding of 
D-branes on curved spaces and a case where both geometric and algebraic 
methods can be applied.

Algebraically, D-branes on a group manifold $G$ are obtained as solutions 
to the gluing conditions of the chiral currents of
the WZW-model \cite{ASa,FFFS,Sta,FS}
\begin{equation} \label{eqAa}
  J(z) = \omega \cdot \bar{J}(\bar z)\,,\qquad 
  \text{for all}\ z=\bar z \,,
\end{equation}
where $\omega$ is an automorphism of the 
Lie algebra $\fg$ of $G$. 
In this paper we will restrict ourselves to the sector $\omega=1$, i.e.\
the so-called symmetry preserving (or untwisted) D-branes.
 
Geometrically, symmetry preserving D-branes wrap conjugacy classes 
$\cC(h) = \{ ghg^{-1} : g\in G \}$.  Even though the conjugacy classes
$\cC(h)$ are contractible within $G$, the D-branes 
correspond to a stable configuration
due to the process of flux stabilization \cite{BDS,MMS,BRS}.
Upon quantization, it is known that the symmetry preserving 
branes are in 1--1
correspondence with the integrable highest weight modules 
of the affine Lie algebra $\wfg$ (at level $k$) 
(see, e.g.\ \cite{BPPZ}) through the
so-called Ishibashi-Cardy boundary states.
In the geometrical picture, the D-brane associated to a boundary 
state labelled by an integrable weight 
$\mu\in P^{(k)}_+$ wraps a conjugacy class 
$\cC(h_\mu)$ where \cite{ASa,Gaa,FFFS,Sta} 
\begin{equation} \label{eqAb}
  h_\mu = e^{\zeta_{\mu}}\,,\qquad
  \zeta_{\mu} = (2\pi i) \frac{\mu+\rho}{k+\dCo} \,.
\end{equation}
[More precisely, the conjugacy classes around which the D-branes 
are wrapped get slightly smeared out (see \cite{ARS,FFFS}).]

In this paper we consider the charge group $C(G,k)$ for 
symmetry preserving D-branes on general compact, connected, 
simply-connected, simple Lie groups $G$.

By considering the process of brane condensation, it 
was argued in \cite{FS} that the charge associated to a D-brane 
labelled by $\lambda\in P^{(k)}_+$ is given by the
dimension $d_\lambda$ of the finite dimensional irreducible
representation $L(\lambda)$ of the horizontal Lie algebra $\fg$,
modulo some integer $x$, i.e.\ the charge group is of the form 
$\ZZ/x\ZZ$, where $x$ is determined such that the  $d_\lambda$
satisfy the fusion rule algebra modulo $x$.  
We find the result that the charge group 
$C(G,k)$, for sufficiently large $k$, is given by $\ZZ/x\ZZ$, where
\begin{equation} \label{eqAc}
 x = \frac{k + \dCo}{\gcd \set{k + \dCo, y}} \,,
\end{equation}
and $y$ is given in Table \ref{tabA}. 
This generalizes the results for $G=SU(N)$ found by, in particular, 
\cite{FS,MMS} (see also \cite{ASa,Stc} for $N=2,3$), in which case 
$y=\lcm(1,2,\ldots,N-1)$ and $\dCo=N$. 
Geometrically, it has been conjectured that the charge group,
in the large volume limit, should be given by the twisted K-theory 
of $G$ \cite{Wi}--\cite{BM}.  
For $SU(N)$ it has been shown that the algebraic and 
geometric definition of the charge group are consistent.  A comparison
for other groups awaits the computation of the relevant twisted K-theory.

The paper is organized as follows.  In Section 2 we review fusion rings,
in particular for the WZW conformal field theory.  In Section 3 we 
present our results.  Section 4 contains some details of the proofs for
the $A_n$ and $C_n$ cases.  In Section 5 we discuss some exceptions 
to the discussion in Section 3, and in Section 6 we discuss symmetry 
properties of the resulting D-brane charges.  We conclude with a discussion
in Section 7.  In the three Appendices we summarize our conventions,
provide a list of generators for the fusion ideals and discuss an alternative
derivation of our results using fusion potentials.

\section{Fusion rings} \label{secfusion}

Consider the chiral algebra $\cA$ of a (rational)
two dimensional conformal field theory together with a (finite)
set of irreducible representations, labelled by $i\in I$ 
(where $I$ is some index set), closed under 
modular transformations of their characters.  To these data 
one can associate a commutative, associative, unital ring $\cF$, the
so-called `fusion ring' of $\cA$.  Explicitly, in terms of a 
(preferred) basis $\{ \phi_i\}_{i\in I}$ of $\cF$ we have
\begin{equation} \label{eqBa}
  \phi_i \times \phi_j = \sum_{k\in I} N_{ij}{}^k\, \phi_k \,,
\end{equation}
where the coefficients $N_{ij}{}^k \in \ZZ_{\geqslant0}$ are called 
the fusion coefficients.  The matrices $N_i$, with components
$(N_i)_j{}^k = N_{ij}{}^k$, are mutually commuting and are
simultaneously diagonalized by the modular matrix $S$.
This leads to the Verlinde formula \cite{Ver}
\begin{equation} \label{eqBb}
  N_{ij}{}^k = \sum_{l\in I} \frac{S_{il} S_{jl} S_{kl}^*}{S_{1l}} \,,
\end{equation}
where $S^*_{ij}$ denotes complex conjugation, and we have denoted
the unit of $\cF$ by $1$.
In terms of $S$, the eigenvalues $\lambda_i^{(l)}$ of $N_i$ 
are given by 
\begin{equation} \label{eqBc}
  \lambda_i^{(l)} = \frac{S^*_{li}}{S_{l1}} \,.
\end{equation}

On abstract grounds (cf.\ \cite{Ge}) it follows that $\cF$ 
is isomorphic to a free polynomial ring factored by 
an ideal $\cI$, i.e.\ 
\begin{equation} \label{eqBd}
  \cF \cong \CC[x_1,\ldots,x_N]/\cI \,.
\end{equation}
In fact, 
when the generators $x_1,\ldots,x_N$ correspond to (a subset of) the primary 
fields $\phi_{i_1} , \ldots , \phi_{i_N}$, 
the ideal $\cI$ can be explicitly characterized as those 
polynomials $P\in \CC[x_1,\ldots,x_N]$ which vanish on 
the set 
\begin{equation*}
 \left\{ \left( \frac{S^*_{l i_1}}{S_{l1}},
 \frac{S^*_{li_2}}{S_{l1}},\ldots,
  \frac{ S^*_{l i_N}}{S_{l1}} \right) \in \CC^N : l\in I
  \right\} \,.
\end{equation*}
We refer to $\cI$ as the `fusion ideal' of $\cF$.
The ideals arising from the fusion rules of a RCFT are actually of 
a special kind, they correspond to so-called Jacobian varieties
(see Appendix \ref{appfuspot}).

In the case of the conformal field theory 
corresponding to the WZW
model for a compact, simple, connected, simply-connected, Lie group $G$,
the relevant representations are the integrable highest weight modules
of the associated (untwisted) affine Lie algebra $\wfg$ at level $k$.
They are parametrized by weights 
$\widehat{\lambda} = (\lambda,k) \in P_+^{(k)}$
(see Appendix \ref{appnot} for a list of our Lie algebra conventions and 
notations).

The modular S-matrix can be expressed as  
\begin{equation} \label{eqBe}
  \frac{S^*_{\mu\lambda}}{S_{\mu1}} = \chi_{\lambda}(\zeta_\mu)\,,
   \qquad  \zeta_\mu = (2\pi i) \frac{\mu+\rho}{k+\dCo}\,,
\end{equation}
where $\chi_\lambda$ is the character of the finite dimensional
irreducible representation $L(\lambda)$ with dominant integral
weight $\lambda$, defined as
\begin{equation}\label{eqBg}
  \chi_\lambda(\sigma) = 
  \frac{\sum_{w\in W} \epsilon(w) e^{(w(\lambda+\rho)|\sigma)}}
  {\sum_{w\in W} \epsilon(w) e^{(w(\rho)|\sigma)}} \,,
\end{equation}
for $\sigma\in \mathfrak{h}^*$. [We identify $\mathfrak{h}^*$ with 
$\mathfrak{h}$, by means of the nondegenerate bilinear form 
$(\ |\ )$, throughout.]

It follows that
\begin{equation}\label{eqBh}
  \chi_\lambda(\zeta_\alpha) \chi_\mu(\zeta_\alpha) = \sum_{\nu\in P_+^{(k)}}
   \  N_{\lambda\mu}{}^\nu\,  \chi_\nu(\zeta_\alpha) \,,
\end{equation}
for all $\alpha\in P_+^{(k)}$.
That is, we can think of the fusion rules of an affine Lie algebra CFT
as truncated tensor products.  [In fact, this leads to a useful 
algorithm for determining explicit fusion rules, the so-called
Kac-Walton algorithm, which is a generalization of 
the Racah-Speiser algorithm 
for determining tensor product coefficients (see, e.g., \cite{DMS}
and references therein).]

Thus we have the following characterization of the fusion ring
for a WZW model at level $k$ 
\begin{equation}
  \cF_k = \ZZ[\chi_1,\ldots,\chi_n]/ \cI_k \,,
\end{equation}
where $\chi_i$ denotes the character of the fundamental 
representation with highest weight $\Lambda_i$, and
where the fusion ideal can be explicitly described 
as those elements in the Grothendieck ring of characters 
which vanish on the all the points $\zeta_\mu$, $\mu\in P^{(k)}_+$.

Elements in $\cI_k$ can easily be constructed by making use of
the identity (see, e.g., \cite{DMS})
\begin{equation} \label{eqBj}
  \chi_{w\cdot\lambda}(\zeta_\mu) = \epsilon(w) 
  \chi_{\lambda}(\zeta_\mu) \,,
\end{equation}
for all $w\in {W}_{\text{aff}}$, where $w\cdot\lambda = 
w(\lambda+\rho)-\rho$
denotes the shifted action of the affine Weyl group (see Appendix
\ref{appnot} for more details).  This implies that 
$\chi_{w\cdot\lambda} - \epsilon(w) \chi_{\lambda} \in
\cI_k$ for all $\lambda\in P_+$ and $w\in {W}_{\text{aff}}$.
In particular, we conclude
that $\chi_\lambda \in \cI_k$ if there exists a positive
root $\alpha\in\Delta_+$ such that $(\lambda+\rho|\alpha) \equiv0
\mod (k+\dCo)$, as this implies $r_\alpha\cdot \lambda = \lambda$,
where $r_\alpha$ denotes the reflection in the root $\alpha$,
and hence $\chi_{\lambda}(\zeta_\mu)=0$, for all $\mu\in P^{(k)}_+$.
For example, we have $\chi_\lambda \in \cI_k$
for all weights $\lambda$ on the hyperplane $(\lambda|\theta)=k+1$.  
Let us call such a weight a `boundary
weight' (at level $k$).  Thus all the characters corresponding to
a level-$k$ boundary weight are in the fusion ideal $\cI_k$.  In fact,
we believe that the fusion ideal $\cI_k$ is generated by the set 
of characters corresponding to the boundary weights 
(this is true for $A_n$ \cite{Ge,GN}). \footnote{With the 
exception of $E_8$ level $2$, see Appendix \ref{appgensets}.} 

While this would characterize the fusion ideal in terms of a finite
set of generators, the number of boundary weights grows rapidly
with $k$ and the corresponding set of generators is still far from 
optimal.  In Appendix \ref{appgensets} we provide what we 
believe is a substantially smaller set of boundary weights, whose
characters generate $\cI_k$.  For $A_n$ and $C_n$ this has been
proven.  The sets provided are not always optimal, but it appears
that what is an optimal set in those cases behaves chaotically
with respect to $k$.

\section{D-brane charge group} \label{secdim}

The charge group for 
symmetry preserving D-branes on group manifolds was investigated
in \cite{Stb,ASb,FSt,FS,MMS}.  In particular, by considering the process 
of D-brane creation and annihilation, it 
was argued in \cite{FS} that the charge of a D-brane 
associated to an integrable weight $\lambda\in P^{(k)}_+$ is given by the
dimension $d_\lambda$ of the finite dimensional irreducible
representation $L(\lambda)$ of the horizontal Lie algebra $\fg$,
modulo some integer $x$.  That is, the charge group is of the form 
$\ZZ/x\ZZ$, where $x$ is determined (as the maximal 
integer) such that the $d_\lambda$
satisfy the fusion rule algebra modulo $x$ \footnote{In fact,
this is a necessary condition.  It is not clear whether this condition
is sufficient in all cases.  Cf.~Section \ref{secsymm}.}
\begin{equation} \label{eqCa}
  d_\lambda d_\mu  = \sum_{\nu\in P_+^{(k)}}
   \  N_{\lambda\mu}{}^\nu\, d_\nu \mod x \,.
\end{equation}
Of course, the dimensions $d_\lambda$ give rise to a homomorphism
\begin{equation} \label{eqCb}
  \text{dim} : \ZZ[\chi_1,\ldots,\chi_n] \to \ZZ\,,\qquad
  \chi_\lambda \mapsto d_\lambda \,,
\end{equation} 
so the problem can be rephrased as the determination 
of the maximal value of $x\in\NN$, such that the dimension
homomorphism \eqref{eqCb} descends to
\begin{equation} 
  \text{dim}' : \cF_k = \ZZ[\chi_1,\ldots,\chi_n] /\cI_k \to \ZZ/x\ZZ \,.
\end{equation}

Clearly, a necessary and sufficient condition for $\text{dim}$ to descend 
to a map on $\cF_k$ is that $\text{dim}(\cI_k) \subset x\ZZ$.  
Note, however, that this analysis presumes a particular presentation
of $\cF_k$ as the quotient of the whole character ring 
$\ZZ[\chi_1,\ldots,\chi_n]$.  
When not all the fundamental representations are integrable at
the level $k$ considered, i.e.\ when $k<\max\{a_i^\vee\}$, the 
fusion ring $\cF_k$
can be presented as a quotient of a subset of the character ring,
and the condition considered above is too strong 
for a homomorphism $\cF_k \to \ZZ/x\ZZ$ to exist.
We will discuss this further in Section \ref{secexcept}, but for now 
we analyze the condition $\text{dim}(\cI_k) \subset x\ZZ$.

It is sufficient to check $\text{dim}(\cI_k) \subset x\ZZ$ on a set
of generators of $\cI_k$.  In particular, assuming the set of boundary
weights generate the fusion ideal, as discussed in Section \ref{secfusion}
we have, 
\begin{equation} \label{eqCc}
  x = \gcd \set{ d_\lambda : \bilin{\lambda}{\theta} = k+1} \,.
\end{equation}

The main result of our paper is that the maximal value of $x$ for
which $\text{dim}(\cI_k) \subset x\ZZ$ is given by the following
formula
\begin{equation} \label{eqCaa}
 x = \frac{k + \dCo}{\gcd \set{k + \dCo , y}} \,,
\end{equation}
where $y$ is an integer, independent of $k$, given in Table \ref{tabA}.

\vspace{3mm}
\begin{center}
\begin{table}[h]
\begin{tabular}{|c|c|c|c|}
\hline
$\fg$ & $\Co$ & $\dCo$ & $y$ \\
\hline
$A_n$ & $n+1$  & $n+1$  & $\lcm \set{1,2,\ldots,n}$    \\
$B_n$ & $2n$   & $2n-1$ & $\lcm \set{1,2,\ldots, 2n-1}$\\
$C_n$ & $2n$   & $n+1$  & $\lcm \set{1,2,\ldots,n,1,3,5,\ldots,2n-1}$ \\
$D_n$ & $2n-2$ & $2n-2$ & $\lcm \set{1,2,\ldots,2n-3}$ \\
$E_6$ & $12$   & $12$   & $\lcm \set{1,2,\ldots,11}$   \\
$E_7$ & $18$   & $18$   & $\lcm \set{1,2,\ldots,17}$   \\
$E_8$ & $30$   & $30$   & $\lcm \set{1,2,\ldots,29}$   \\
$F_4$ & $12$   & $9$    & $\lcm \set{1,2,\ldots,11}$   \\
$G_2$ & $6$    & $4$    & $\lcm \set{1,2,\ldots,5}$    \\
\hline
\end{tabular}\vspace{3mm}
\caption{The Coxeter number $\Co$ and dual Coxeter number $\dCo$ for each
(finite dimensional) simple Lie algebra, 
and the number $y$ occurring in Eqn.~\eqref{eqCaa}.}
\label{tabA}
\end{table}
\end{center}

This generalizes a result in \cite{FS} (see also \cite{MMS}) 
for $\fg = A_n$.  
Note that in all cases we can write $y = \lcm \set{y_{\alpha}}$, for
some integers $y_{\alpha}$.  For the simply laced Lie algebras, 
these integers are found to be consecutive, from $1$ up to $\dCo-1$, but for
non-simply laced Lie algebras, additional integers are required.  Note,
moreover, that for all Lie algebras, with the exception of $C_n$, $y$
is given by the lowest common multiple of all integers from 
$1$ up to $\Co-1$ (cf.\ the discussion in Section \ref{secsymm}).
 
We will prove Eqn.~\eqref{eqCaa}
in Section \ref{secanalres} for the
algebras $A_n$ and $C_n$.  In the other cases we have strong numerical
evidence, based on explicit computations for levels up to $k=5000$,
using both the explicit fusion rules (calculated with the help
of \cite{Sch}) and sets of elements in the ideal constructed with
the procedure outlined in Section \ref{secfusion}.
In Appendix \ref{appgensets} we provide what we 
believe is a set of generators for the ideal $\cI_k$.

For an intuitive way to see how the formula for $x$ may arise, 
let us consider the Weyl dimension formula
\begin{equation} \label{eqWeyldim}
  d_\lambda = \prod_{\alpha \in \Delta_+} 
  \frac{\bilin{\lambda + \rho}{\alpha}}{\bilin{\rho}{\alpha}} \,.
\end{equation}
When $\lambda$ is a boundary weight, we find that $(\lambda+\rho|\theta) $
contributes a factor $k + \dCo$ to the numerator, so it is not 
surprising that it appears in the formula for the greatest common 
divisor of these dimensions.

Let us now concentrate on the factors $\bilin{\rho}{\alpha}$ of the
denominator.  For simply laced Lie algebras these factors run from $1$ up to
$\bilin{\rho}{\theta} = \dCo - 1$ (with some repetitions).  These
are the $y_{\alpha}$ whose least common multiple gives the parameter
$y$ in \eqref{eqCaa}.  For non-simply laced algebras, 
the factors $\bilin{\rho}{\alpha}$ need not
be integers.  It is easily verified that setting
\begin{equation}
y_{\alpha} = 
\begin{cases}
  \bilin{\rho}{\alpha} & \text{if } \bilin{\rho}{\alpha} \in \ZZ\,, \\
  \bilin{\rho}{\alpha^{\lor}} & \text{if } \bilin{\rho}{\alpha} \notin 
  \ZZ\,,
\end{cases}
\end{equation}
and taking $y = \lcm \set{y_{\alpha}: \alpha\in\Delta_+}$, reproduces the 
results for $y$ given in Table \ref{tabA}.

\section{Analytical Results} \label{secanalres}

In this section we provide explicit proofs of 
\eqref{eqCaa} for the algebras $A_n$ and
$C_n$.  While the result for $A_n$ has already been established in
\cite{FS,MMS}, using simple currents and outer automorphisms, respectively,
we will present an alternative proof which 
generalizes, at least in principle, to the other simple Lie algebras.

\subsection{$\boldsymbol{A_n}$} \label{secA_n}

We claim (see \eqref{eqCaa}) that
for symmetry preserving D-branes defined on the group $A_n$, 
($n\geqslant 1$), at levels $k \geqslant 1$, the charge group has the form
$\ZZ/x\ZZ$, where
\begin{equation} \label{eqDa}
  x = \frac{k+n+1}{\gcd \set{k+n+1, \lcm \set{1,2,\ldots,n}}} \,.
\end{equation}

The first proof we present is based on the results in \cite{FS}.
There it is shown by an induction argument, using the simple currents
of $A_n$, that the fusion ideal is generated by the 
characters of the boundary modules with weights $\{ k\Lambda_1 +
\Lambda_i : i=1,\ldots,n\}$.  We thus have 
\begin{equation} \label{eqDb}
  x = \gcd \set{ d_{k \Lambda_1 + \Lambda_i} : 
  i = 1,2,\ldots,n}\,.
\end{equation}
To prove that this equals \eqref{eqDa} we proceed as follows:
The dimensions $d_{k \Lambda_1 + \Lambda_i}$ are given explicitly by
\begin{equation} \label{eqDc}
  d_{k \Lambda_1 + \Lambda_i} = \frac{i}{k+i} 
  \binom{n+1}{i} \binom{k+n+1}{k} = \binom{k+i-1}{k}
  \binom{k+n+1}{n+1-i} \,.
\end{equation}
The second expression immediately shows that
\begin{equation}\label{eqDca}
    \gcd \set{ d_{k \Lambda_1 + \Lambda_i} : i = 1, 
   \ldots,n} \geqslant \gcd \set{\binom{k+n+1}{i} : i = 1 , \ldots , n} \,.
\end{equation}
To show the reverse inequality, we prove the following relation
\begin{equation} \label{eqDd}
  \sum_{i=1}^{n+1-j} (-1)^{i-1} \binom{n+1-i}{j} d_{k
  \Lambda_1 + \Lambda_i} = \binom{k+n+1}{j} \,.
\end{equation}
Substituting the first expression in \eqref{eqDc} 
for $d_{k \Lambda_1 + \Lambda_i}$ into \eqref{eqDd}, 
we find that this relation reduces to
\begin{equation*}
  \binom{k+M}{k} \sum_{i=0}^M (-1)^{i-1} \frac{i}{k+i} \binom{M}{i} = 1\,,
\end{equation*}
where we have put $M = n + 1 - j$.  
Using $\frac{i}{k+i} = 1 - \frac{k}{k+i}$ 
and $\sum_{i=0}^M (-1)^i \binom{M}{i}=0$, 
we reduce this further, obtaining
\begin{equation*}
  k \binom{k+M}{k} \sum_{i=0}^M (-1)^i \frac{1}{k+i} \binom{M}{i} = 1\,.
\end{equation*}
This equation is clearly true for $M = 0$ and all $k \in
\ZZ_+$, so we can finish the proof of the relation by induction
on $M$ (for all $k$).  In the induction step we use 
\begin{equation} \label{eqDf}
  \binom{M+1}{i} = \binom{M}{i} + \binom{M}{i-1} \,,
\end{equation} 
and change the summation variable in
the sum arising from the second term from $i$ to $i+1$.  This then
establishes Eqn.~\eqref{eqDd} and, hence, the equality sign in 
Eqn.~\eqref{eqDca}.
Finally, it is proven in Appendix C of \cite{MMS} that 
\begin{equation} \label{eqMalAppC}
  \gcd \set{\binom{k+n+1}{i} : i =1,\ldots,n} = 
  \frac{k+n+1}{\gcd \set{k+n+1, \lcm \set{1,2,\ldots,n}}} \,,
\end{equation}
completing the proof of Eqn.~\eqref{eqDa}.

It is clear from the discussion in Section \ref{secfusion} that the
charge group parameter $x$ may be obtained from any set of generators
of the fusion ideal.  It is obviously highly desirable to find
generators for which the above evaluation gives nice expressions.  The
proof given above uses the generators corresponding to the weights
$\set{k \Lambda_1 + \Lambda_i : i=1,\ldots,n}$ whose dimensions are a little
cumbersome to manipulate -- hence the proof is not straightforward.
An alternative set of generators for the fusion ideal was 
constructed in \cite{Ge} (see also \cite{MRS}), and given by 
\begin{equation}  \label{eqDe}
  \set{\func{}{k+i} \Lambda_1 : i = 1,\ldots,n} \,,
\end{equation}
This set is particularly nice because
\begin{equation}
 d_{(k+i) \Lambda_1} = \binom{k+i+n}{n}   \,. 
\end{equation}
We now have 
\begin{align} \label{eqDAab}
  x & = \gcd \set{\binom{k+n+1}{n} , \binom{k+n+2}{n},\ldots, 
  \binom{k+2n}{n}} \\
  & = \gcd \set{\binom{k+n+1}{1} , \binom{k+n+2}{2},\ldots, 
  \binom{k+2n}{n}} \nonumber \\
  & = \gcd \set{\binom{k+n+1}{1} , \binom{k+n+1}{2},\ldots, 
  \binom{k+n+1}{n}} \nonumber \,,
\end{align}
by repeatedly using relation \eqref{eqDf}.
With Eqn.~\eqref{eqMalAppC}, this completes the proof of \eqref{eqDa}.
Another reason for why the set of generators \eqref{eqDe} is nice
is that this set is intimately related to the fusion
potential for $A_n$ \cite{Ge} (see Appendix \ref{appfuspot}).

\subsection{$\boldsymbol{C_n}$} \label{secC_n}

For $C_n$ ($n\geqslant 2$), at levels $k \geqslant 1$, 
the result \eqref{eqCaa} is explicitly given by
\begin{equation} \label{eqDBa}
  x = \frac{k+n+1}{\gcd \set{k+n+1, \lcm \set{1,2, 
  \ldots,n,1,3,5,\ldots,2n-1}}} \,.
\end{equation}
Consider, as in the second $A_n$ proof, the set of weights
$\set{\func{}{k+i} \Lambda_1 : i =1,\ldots,n}$.  
While, contrary to the $A_n$ case, these weights in general 
do not correspond
to elements of the fusion ideal, we can use them to construct 
elements in the ideal by using the (shifted action of the) 
affine Weyl reflection $r_0$, as explained in Section \ref{secfusion}.  
We find that
\begin{equation}
  r_0 \cdot \func{}{k+i} \Lambda_1 = \func{}{k+2-i} \Lambda_1 \,.
\end{equation}
Hence, the following set of combinations of 
characters are in the fusion ideal $\cI_k$ \footnote{At low levels,
some of these weights will lie outside the fundamental Weyl chamber,
and have to be reflected back to the fundamental Weyl chamber.
This does however not impact on the general validity of the analysis
below since the dimensions $d_\lambda$ are invariant under 
the (shifted) action of $W$.}
\begin{equation}
  \{ \chi_{(k+1)\Lambda_1}, \chi_{(k+2)\Lambda_1}+\chi_{k\Lambda_1},
  \chi_{(k+2)\Lambda_1}+\chi_{(k-1)\Lambda_1}, \ldots,
  \chi_{(k+n)\Lambda_1}+\chi_{(k+2-n)\Lambda_1} \} \,.
\end{equation}
In fact, as shown in \cite{MRS,BMRS,GS}, these combinations actually
constitute a set of generators of $\cI_k$.

Using 
\begin{equation}
  d_{k \Lambda_1} = \binom{k+2n-1}{2n-1} \,,
\end{equation}
we find
\begin{multline} \label{eqDBbb}
  x = \gcd \left\{ \binom{k+2n}{2n-1} , \binom{k+2n+1}{2n-1} + 
   \binom{k+2n-1}{2n-1} \,, \right. \\
\left. \binom{k+2n+2}{2n-1} + \binom{k+2n-2}{2n-1}, \ldots , 
\binom{k+3n-1}{2n-1} + \binom{k+n+1}{2n-1} \right\} \,.
\end{multline}
Upon judicious use of the identity \eqref{eqDf}
this can be rewritten as 
\begin{equation}
  x = \gcd \set{ \binom{k+n+1}{1}, \binom{k+n+2}{3},
  \ldots,\binom{k+2n-1}{2n-3},\binom{k+2n}{2n-1}} \,.
\end{equation}
Using a generalisation of Eqn.~\eqref{eqMalAppC}, which may
be proven the same way, we arrive at Eqn.~\eqref{eqDBa}.

Alternatively, one may use a different set of generators of $\cI_k$,
corresponding to the boundary weights $\{ k\Lambda_1 + \Lambda_i :
i=1,\ldots,n\}$.  This leads to the same result, and we leave the details 
to the reader.

\section{Exceptions} \label{secexcept}

In Section \ref{secdim} we have argued that a necessary and sufficient 
condition for the dimension homomorphism \eqref{eqCb} on the character 
ring $\ZZ[\chi_1,\ldots,\chi_n]$ to descend to $\cF_k$ is that 
$\text{dim}(\cI_k) \subset x\ZZ$.  Let us illustrate 
that this condition is sometimes too strong for a homomorphism
$\cF_k \to \ZZ/x\ZZ$ to exist,
i.e.\ does not give the charge group as defined by 
Eqn.\ \eqref{eqCa}, by looking, for example,
at the Lie algebra $G_2$ at level $k=1$.  

There are two integrable representations, corresponding to the weights
$\Lambda_0$ (the unit of the fusion
ring) and $\Lambda_2$. The nontrivial fusion rule reads
\begin{equation} \label{eqG2fus}
  \phi_{\Lambda_2} \times \phi_{\Lambda_2} = \phi_{\Lambda_0} + 
  \phi_{\Lambda_2} \,.
\end{equation}
Substituting the appropriate dimensions
we find that this equation is satisfied modulo $x =
41$.  However, \eqref{eqCc} would suggest $x=1$, as
the representations with boundary weights $\Lambda_1$ and 
$2 \Lambda_2$ have dimensions $14$ and $27$, respectively.

To explain what is going on, consider the tensor product decomposition 
\begin{equation} \label{eqG2tensprod}
  \chi_{\Lambda_2}\chi_{\Lambda_2} = \chi_{\Lambda_0}  + 
  \chi_{\Lambda_2} + \chi_{\Lambda_1} + \chi_{2\Lambda_2} \,.
\end{equation}
The truncation in going from the tensor product rule to Eqn.\ 
\eqref{eqG2fus} therefore consists of setting $\chi_{\Lambda_1} + 
\chi_{2\Lambda_2}$
to zero rather than setting $\chi_{\Lambda_1}$ and
$\chi_{2\Lambda_2}$ to zero separately, as we would have required in
Section \ref{secdim}.

The resolution of this apparent dilemma lies of course in the 
fact that the fundamental representation with weight $\Lambda_1$ 
is not integrable at level $k=1$, so rather than writing the 
fusion ring as 
\begin{equation}
  \cF_k = \ZZ[\chi_1 , \chi_2 ]/\cI_k \,,
\end{equation}
with $\cI_k = \langle \chi_1, \chi_2^2 -\chi_1 - \chi_2 - 1 \rangle$
we can write 
\begin{equation}
  \cF_k = \ZZ[\chi_2 ]/\cI'_k \,,
\end{equation}
with $\cI'_k = \langle \chi_2^2 - \chi_2 - 1 \rangle$, and clearly
$\text{dim}(\cI_k') \subset 41\ZZ$.  Thus this is the correct 
presentation to use for determining the charge group in this case.

Summarizing, when not all the fundamental representations are integrable 
at the level $k$ one is considering, i.e.\ $k < \max \set{a_i^\vee}$,
the presentation of the fusion ring as a quotient of the complete
character ring is redundant, as it introduces artificial generators 
(the non-integrable fundamental representations) which must then be 
factored out again.
In those cases it is a simple matter to compute $x$ directly from 
\eqref{eqCa}, using the explicit fusion coefficients (which
were calculated with the help of \cite{Sch}).
We have tabulated these values in 
Table \ref{tabB} whenever they disagree with the 
result of \eqref{eqCaa}.  Note that in many cases in which 
$k < \max \set{a_i^\vee}$, formula \eqref{eqCaa} continues to hold.
[Note also that for $E_8$ at level $1$ the charge group is just
$\ZZ$ as there is only the trivial fusion rule.]

\vspace{3mm}
\begin{center}
\begin{table}[h] 
\begin{tabular}{|c|c|c|}
\hline
$\mathfrak{g}$ & $k$ & $x$ \\
\hline
$B_n$ & $1$ & $2 \gcd \func{}{2n\func{}{n+1} , n2^n , 2^{2n-1}-n-1}$ \\
$D_n$ & $1$ & $\gcd \func{}{2n-1 , 2^{2n-3}-n , 2^{2n-2}-1}$ \\
$E_6$ & $1$ & $26$ \\
$E_7$ & $1$ & $3135$ \\
$E_8$ & $1$ & $\infty$ \\
$E_8$ & $2$ & $4$ \\
$F_4$ & $1$ & $649$ \\
$G_2$ & $1$ & $41$ \\
\hline
\end{tabular}\vspace{3mm}
\caption{Exceptional cases}
\label{tabB}
\end{table}
\end{center}

Finally, note that in almost all exceptional cases $x$ is not
a divisor of $k+\dCo$.  For reasons to be discussed in Section 
\ref{secconcl} these exceptional results therefore are not
consistent with the prediction that would follow from a computation
of the twisted K-theory.

\section{Symmetry considerations} \label{secsymm}

In this section we discuss some of the symmetry properties of the
charges $D_\lambda = d_{\lambda} \mod x$, with $x$ given by 
\eqref{eqCaa}.  
To simplify the discussion
let us extend the dimension formula for $d_\lambda$ in
\eqref{eqWeyldim} to all (not necessarily dominant) weights $\lambda$.
It is obvious that $d_\lambda$ (and hence $D_\lambda$) is invariant
under both the Weyl group $W$ and the group of outer
automorphisms $\text{Aut}^0(\fg)$ of $\fg$.  
For elements $\Omega\in \text{Aut}^0(\wfg)$, i.e.\ symmetries
of the affine Dynkin diagram, we find however that 
\begin{equation} \label{eqFa}
  D_{\Omega \lambda} = e^{ 2\pi i (\Lambda_i|\rho)} D_\lambda \,,
\end{equation}
where $i$ is determined by $\wLa_i = \Omega \wLa_0$.  
We recall that 
\begin{equation} \label{eqFb}
  e^{ 2\pi i (\Lambda_i|\rho)} = \epsilon(w_\Omega) \,,
\end{equation}
where $w_\Omega\in W$ is the Weyl group element determined by 
(see \cite{DMS} for details)
\begin{equation} \label{eqFc}
  \Omega\widehat\lambda = k (\Omega-1)\wLa_0 + w_\Omega \widehat\lambda \,.
\end{equation}
This symmetry reflects a similar symmetry in the fusion rule coefficients
\begin{align} \label{eqFd}
  N_{\Omega(\lambda) \mu}{}^{\Omega(\nu)}  & = N_{\lambda\mu}{}^\nu \,,
  \nonumber \\
  N_{\Omega(\lambda) \mu}{}^\nu & = N_{\lambda \Omega(\mu)}{}^\nu \,.
\end{align}

It is well-known that $\text{Aut}^0(\wfg)/\text{Aut}^0(\fg)$ is isomorphic to
the center $Z(\fg)$ of $\fg$.  The homomorphism for $\Omega\in 
\text{Aut}^0(\wfg)$, such that $\Omega \wLa_0 = \wLa_i$,
is explicitly given by
\begin{equation} \label{eqFe}
  \Omega \mapsto z_i = e^{ -2\pi i \Lambda_i \cdot H } \quad \in Z(\fg) \,.
\end{equation}
Geometrically, acting with an element of the center corresponds to
rotating the conjugacy class around which the D-brane is wrapped 
and hence should leave invariant the charge, up to a possible sign
\cite{MMS,Stc}.  In fact, in \cite{MMS} (see also \cite{Stc,Std} 
for $SU(3)$, where this was referred to as the `multiplet structure') 
it was shown that requiring the invariance \eqref{eqFa}, 
in the case of $SU(N)$, leads to the same charge group.  This
is however not the case for the other Lie groups.

Finally, the charges $D_\lambda$ have a symmetry under the affine 
Weyl group $W_{\text{aff}} = T(Q^\vee) \rtimes W$, that reflects 
the symmetry in, e.g., the 
modular S-matrices and specialized characters (see \eqref{eqBj})
\begin{equation} \label{eqFf}
  D_{\hat w\cdot \lambda} = \epsilon(\hat w) D_\lambda \,, 
  \qquad \hat w \in W_{\text{aff}} \,.
\end{equation}
The affine Weyl group $W_{\text{aff}}$ and the outer automorphism
group $\text{Aut}^0(\wfg)$ combine into a group 
which projects onto the finite weight space as 
$T(Q^*) \rtimes W$, where $Q^*$ is the lattice dual to $Q$.
It turns out that, apart from a few low level exceptions,
enforcing charge invariance \eqref{eqFf} under $T(Q^*) \rtimes W$
determines the charge group in all cases.  Including the expected
invariance under affine outer automorphisms is therefore not in
conflict with the charge group results.

Rather surprisingly, when studying the action of the extended affine
Weyl group $W_{\text{aff}}' = T(P) \rtimes W$ on $D_\lambda$, where
$P$ is the weight lattice of $\fg$ (in the simply laced case $P =
Q^*$), one finds that \eqref{eqFf} holds for $\hat w
\in W_{\text{aff}}'$ as well, in all cases except $C_n$, $n$ not
a power of $2$, at certain levels.  

\begin{figure}[h]
\begin{center}
\includegraphics[height=10cm]{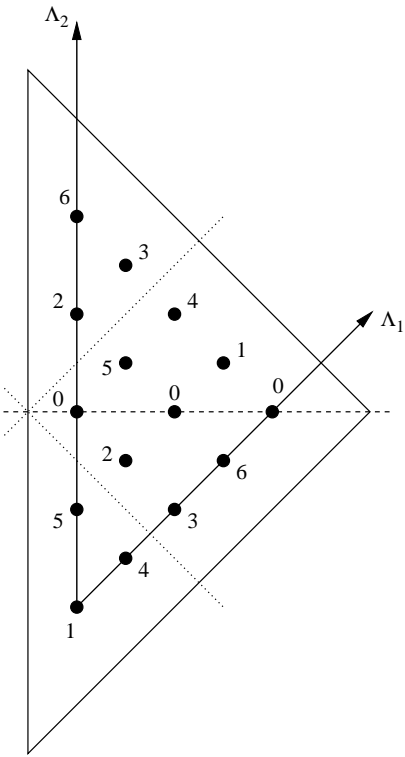}
\end{center}
\caption{The integrable highest weights of $C_2$ at level $4$
(shifted by $\rho$) and the charges of
the corresponding symmetry preserving D-branes.  The charges take values
mod $7$.  Note that the charges are negated (mod $7$) under reflection
about the dashed and dotted lines.} \label{figC24charges}
\end{figure}

In Figure
\ref{figC24charges}, we show as an example, the charges for $C_2$ at
level $4$ (for which \eqref{eqFf} does hold for $\hat w \in
W_{\text{aff}}'$).  The symmetry about the dashed line corresponds to
the outer automorphism of $C_n^{(1)}$.  It is easily seen to be an
element of $W$ of negative sign composed with a translation by
$\func{}{k + \dCo} \Lambda_2 = 7 \Lambda_2$ (note $\Lambda_2 \in
Q^*$).  The dotted lines correspond to more general symmetries.  The
lower line is an element of $W$ of negative sign composed with a
translation by $7 \Lambda_1$ (but $\Lambda_1 \notin Q^*$).  The upper
dotted line corresponds to the composition of an element of $W$ of
negative sign composed with a translation by $7 \left( \Lambda_1 +
\Lambda_2 \right)$.

Requiring \eqref{eqFf} to hold for all 
$\hat w \in W_{\text{aff}}'$ for $C_n$ would lead to the requirement
that $x$ is given by \eqref{eqCaa} with 
$y = \lcm\{1,2,\ldots,2n-1\} = \lcm\{1,2,\ldots,\Co-1\}$, and thus 
would lead to a universal expression for all compact, simple, connected,
simply-connected Lie groups (cf.\ the discussion in Section \ref{secdim}).
This is suggestive of the idea that maybe, in the case of
$C_n$, the fusion rule constraints \eqref{eqCa}
on the charge group do not take into account all physical constraints
that the D-brane charge group needs to satisfy.  For this reason it would 
be very interesting to compute the twisted K-theory for $C_n$ and make a 
comparison.

\section{Conclusions and discussion} \label{secconcl}

It has been argued \cite{Wi}-\cite{BM} that the charges of 
D-branes on a manifold $X$ in the background of a nontrivial 
NS-NS $B$-field, i.e.\ an element $[H]\in H^3(X,\ZZ)$
(where $H=dB$ locally),
are classified by twisted K-theory groups \cite{Ros},
at least in the large volume limit. 
In particular, for $D$-branes on a compact,
simple, simply connected group manifold $G$, by
$K^{\bullet}(G, (k+\dCo)[H_0])$, where $[H_0]$ is the generator of
$H^3(G,\ZZ) \cong \ZZ$ and $k$ is the level of the 
(supersymmetric) WZW model \cite{FS}.  
It has been verified that this proposal is consistent with
the results based on boundary conformal field theory, i.e.\
the techniques used in this paper, for the 
the cases $G=SU(2)$ \cite{ASa,FS}, $G=SU(3)$
\cite{FS,Stc} and $G=SU(N)$ \cite{FS,MMS}. 

In other cases, the comparison of the results of this paper
with the corresponding twisted K-theory awaits a computation
of the latter.  In particular this would settle the issue of
whether the consistency equations of \cite{FS}, used in 
this paper, suffice to determine the charge group, in particular in
the case of $C_n$ (as discussed in Section \ref{secsymm}).

Beyond that, it would be very interesting to have
a direct understanding of why the two approaches should lead to the 
same result.  For instance, while it is obvious in twisted K-theory
that $x$ is a divisor of $k+\dCo$ (through the
Atiyah-Hirzebruch spectral sequence \cite{Ros}), there does not
appear to be a simple explanation for this in our approach.

By a result of Freed, Hopkins and Telemann (cf.\ \cite{Fre}) the fusion 
algebra $\cF_k$ for the level-$k$ WZW model can be identified with 
the twisted equivariant K-theory $K_G(G,(k+\dCo)[H_0])$.  
It would be very interesting to establish a direct link
between the equivariant and non-equivariant
twisted K-theory of $G$ on the one hand, and the relation 
between the fusion algebra 
$\cF_k$ and homomorphisms $\cF_k \to \ZZ/x\ZZ$ on the other.

A particularly concrete presentation of twisted K-theory is 
as the Grothendieck K-theory of (isomorphism classes of) bundle
gerbe modules \cite{BCMMS}.  Bundle gerbes show up naturally in
the discussion of the WZW model \cite{GR,Mic}, and the bundle gerbe 
for $G=SU(N)$ has recently been constructed in \cite{GR,Mei}.
It would be interesting to explicitly construct the bundle gerbe 
modules corresponding to a particular D-brane charge.

Finally, it would be interesting to extend the results of this paper 
to the non-symmetry preserving branes (i.e.\ twisted branes) 
\cite{FFFS,Stc,AFQS},
branes on coset spaces \cite{Gab,MMSa}--\cite{FSb}, and beyond \cite{Ste,RSc}.
We hope to report on this in the future.

\appendix

\section{Lie algebra notations} \label{appnot}

We summarize the Lie algebra notations used in this paper for convenience. 
Let $\fg$ denote a simple finite-dimensional Lie algebra
of rank $n$, and $\wfg$ 
its untwisted affine extension, i.e.\ $\wfg = (\fg \otimes \CC[t,t^{-1}])
\oplus \CC k$.  The Cartan subalgebra of $\fg$ is denoted by 
$\mathfrak{h}$ and its dual by $\mathfrak{h}^*$.
The set of positive roots in $\mathfrak{h}^*$ is denoted by $\Delta_+$, 
the set of simple roots by $\Pi = \{ \alpha_1,\ldots,\alpha_n\}$.
The fundamental weights $\Lambda_i$ of $\fg$ are defined by 
\begin{equation} \label{eqAppAa}
  (\Lambda_i|\alpha_j^\vee) = \delta_{ij} \,,
\end{equation}
where $\alpha^\vee = 2\alpha/(\alpha|\alpha)$ is the co-root corresponding
to $\alpha$.  Here we have normalized 
the nondegenerate positive bilinear form $(\ |\ )$ on $\mathfrak{h}^*$ 
such that $(\theta|\theta)=2$, where $\theta$ is the 
highest root of $\fg$.

The Weyl vector is denoted by $\rho$ and is defined by
\begin{equation}\label{eqAppAb}
  (\rho|\alpha_i^\vee) = 1 \,,
\end{equation}
for all $i=1,\ldots,n$, i.e., $\rho = \frac12 
\sum_{\alpha\in\Delta_+} \alpha=\sum_{i=1}^n \Lambda_i$.
The marks $\{a_i\}_{i=1}^n$ and 
co-marks $\{a_i^\vee\}_{i=1}^n$ are given by
\begin{equation}\label{eqAppAc}
  \theta = \sum_{i=1}^n a_i \alpha_i = 
  \sum_{i=1}^n a_i^\vee \alpha_i^\vee \,.
\end{equation}
We denote the Coxeter and dual Coxeter numbers by $\Co$ and $\dCo$, 
respectively (see Table \ref{tabA}),
the root and co-root lattices by $Q$ and $Q^\vee$,
and the lattices of integral weights and dominant integral
weights by $P$ and $P_+$, and
integrable weights at level $k$ by $P_+^{(k)} = \{
\lambda \in P_+ : (\lambda|\theta)\leqslant k\}$.

For the affine Lie algebra $\wfg$ we use the same notations except
that we put a hat on the symbols.  To simplify notations, however,
we will sometimes simply use the notation $\lambda$ for an
integrable weight $\hat\lambda=(\lambda,k)$, when the level $k$
is understood.

We denote the Weyl groups of $\fg$ and $\wfg$ by $W$ and $\widehat W$,
respectively, and the shifted Weyl group action by a dot, i.e.\ 
\begin{equation}\label{eqAppAd}
  w\cdot \lambda = w(\lambda+\rho) - \rho \,,
\end{equation}
and similarly for $\hat w \in \widehat W$.  
The determinant of a Weyl group element $w$ is denoted by
$\epsilon(w)$.  The
projection of $\widehat W$ onto $\mathfrak{h}^*$ is given by
\begin{equation}\label{eqAppAe}
  W_{\text{aff}} = T(Q^\vee) \rtimes W \,,
\end{equation}
where $T(\Gamma)$ denotes the translation group of the lattice $\Gamma$,
and the semi-direct product structure is given by the relation 
$w t_\alpha w^{-1} = t_{w\alpha}$, i.e.\ $(t_\gamma,w) \circ 
(t_{\gamma'},w') = (t_\gamma t_{w\gamma'}, ww')$.
In particular, $t_{\theta^\vee} = r_0 r_\theta$.
Explicitly, the shifted action of 
$t_\gamma \in W_{\text{aff}}\,,\ \gamma\in Q^\vee$, is given by 
\begin{equation}\label{eqAppAf}
  t_\gamma \cdot \lambda = \lambda + (k+\dCo) \gamma \,.
\end{equation}

\section{Generating Subsets} \label{appgensets}

In this appendix we provide a subset of the boundary weights,
i.e.\ weights satisfying $\bilin{\lambda}{\theta} = k+1$,
which we believe form a generating set 
for the fusions ideals $\cI_k$.  For $A_n$ and $C_n$ this has been 
proven, in the other cases we tested numerically for $k <
5000$ that these sets give rise to the result \eqref{eqCaa}.

When $k$ is sufficiently small, some of the weights given
below will not lie in the fundamental chamber -- these weights should
be discarded to get a generating set, except in the following cases
where it is additionally required to include another weight.  This
occurs with $E_7$ at levels $2$ and $3$, adding
$\func{}{0,0,0,1,0,0,0}$ and $\func{}{0,0,0,0,0,0,2}$ respectively,
and $E_8$ at levels $4$ and $12$, adding respectively
$\func{}{0,1,0,0,0,0,1,0}$ and $\func{}{0,2,1,0,0,0,0,1}$.  However,
it may be verified that for $E_8$ at level $2$, there is no set of
weights in the fundamental chamber whose dimensions have the greatest
common divisor given in \eqref{eqCaa}.  
In this case it is necessary to include weights outside
the fundamental chamber (extending the dimension formula in the
obvious way). 

Fundamental weights are denoted by $\Lambda_i$.  For the exceptional
algebras, we have listed the weights appearing in the generating sets
in terms of their Dynkin labels, for clarity, using the conventions
of \cite{DMS}.

\begin{description}
\item[$\boldsymbol{A_n}$]
\begin{equation*}
\left\{ k\Lambda_1 + \Lambda_i : i=1, \ldots, n\right\} \,.
\end{equation*}
\item[$\boldsymbol{B_n}$]
\begin{equation*}
\set{\func{}{k+1} \Lambda_1, \func{}{k-1} \Lambda_1 + 2 \Lambda_n} 
\cup \set{\func{}{k-1} \Lambda_1 + \Lambda_i, \func{}{k-2} \Lambda_1 + 
\Lambda_i + \Lambda_n : i = 2 , \ldots , n-1} \,.
\end{equation*}
\item[$\boldsymbol{C_n}$]
\begin{equation*}
\left\{ k\Lambda_1 + \Lambda_i : i=1, \ldots, n\right\} \,.
\end{equation*}
\item[$\boldsymbol{D_n}$]
\begin{multline*}
\left\{k \Lambda_1 + \Lambda_n, \func{}{k-1} \Lambda_1 + 2 \Lambda_n, 
 \func{}{k-1} \Lambda_1 + \Lambda_{n-1} + 
\Lambda_n, \func{}{k-3} \Lambda_1 + \Lambda_{n-1} + 3 \Lambda_n \right\} \\
\cup \set{\func{}{k-1} \Lambda_1 + \Lambda_i, \func{}{k-3} \Lambda_1 + 
\Lambda_i + 2 \Lambda_n : i = 2 , \ldots , n-2} \\
\cup \set{\func{}{k-3} \Lambda_1 + \Lambda_i + \Lambda_j : 2 
\leqslant i < j \leqslant n-2} \,.
\end{multline*}
\item[$\boldsymbol{E_6}$]  For $k$ odd, 
\begin{multline*}
\{ \func{}{0,0,0,0,0,\tfrac{k+1}{2}} ,
\func{}{0,0,0,0,2,\tfrac{k-1}{2}} , 
\func{}{1,0,0,0,1,\tfrac{k-1}{2}} , \func{}{0,1,0,1,0,\tfrac{k-3}{2}} , \\
\func{}{0,0,0,0,6,\tfrac{k-5}{2}} , \func{}{0,0,0,3,0,\tfrac{k-5}{2}} ,
\func{}{3,0,0,0,3,\tfrac{k-5}{2}} , \func{}{1,0,1,0,4,\tfrac{k-7}{2}} \}\,.
\end{multline*}
For $k$ even,
\begin{multline*}
\{ \func{}{0,0,0,0,1,\tfrac{k}{2}} , \func{}{0,0,0,0,3,\tfrac{k-2}{2}}
, 
\func{}{0,0,0,1,1,\tfrac{k-2}{2}} , \\
\func{}{0,0,1,0,0,\tfrac{k-2}{2}} , \func{}{0,2,0,0,1,\tfrac{k-4}{2}}
, 
\func{}{1,0,1,0,1,\tfrac{k-4}{2}} , \\
\func{}{0,1,0,0,5,\tfrac{k-6}{2}} , \func{}{0,1,0,2,1,\tfrac{k-6}{2}}
, 
\func{}{2,0,0,1,3,\tfrac{k-6}{2}} \} \,.
\end{multline*}
\item[$\boldsymbol{E_7}$]  For $k$ odd,
\begin{multline*}
\{ \func{}{\tfrac{k+1}{2},0,0,0,0,0,0} , 
\func{}{\tfrac{k-1}{2},0,0,0,1,0,0} , \\
\func{}{\tfrac{k-3}{2},0,1,0,0,0,0} , 
\func{}{\tfrac{k-5}{2},0,0,2,0,0,0} , 
\func{}{\tfrac{k-7}{2},0,0,0,0,8,0} , \\
\func{}{\tfrac{k-7}{2},0,0,0,4,0,0} , 
\func{}{\tfrac{k-7}{2},0,1,0,2,0,0} , 
\func{}{\tfrac{k-9}{2},0,0,2,0,4,0} , \\
 \func{}{\tfrac{k-9}{2},0,1,0,0,6,0} , 
\func{}{\tfrac{k-9}{2},0,1,0,2,2,0} \} \,.
\end{multline*}
For $k$ even,
\begin{multline*}
\{ \func{}{\tfrac{k}{2},0,0,0,0,1,0} , 
\func{}{\tfrac{k-2}{2},1,0,0,0,0,0} , \\
\func{}{\tfrac{k-2}{2},0,0,0,0,1,1} , 
\func{}{\tfrac{k-4}{2},1,0,0,0,2,0} , 
\func{}{\tfrac{k-4}{2},1,0,0,1,0,0} , \\
\func{}{\tfrac{k-4}{2},0,0,1,0,0,1} , 
\func{}{\tfrac{k-6}{2},1,0,1,0,1,0} , 
\func{}{\tfrac{k-6}{2},0,0,1,1,0,1} , \\
\func{}{\tfrac{k-8}{2},1,0,0,3,0,0} , 
\func{}{\tfrac{k-8}{2},0,0,0,0,7,1} , 
\func{}{\tfrac{k-8}{2},0,0,0,3,1,1} , \\
 \func{}{\tfrac{k-10}{2},1,0,1,0,5,0} , 
\func{}{\tfrac{k-10}{2},1,0,1,1,3,0} \} \,.
\end{multline*}
\item[$\boldsymbol{E_8}$]  For $k$ odd,
\begin{multline*}
\{ \func{}{\tfrac{k+1}{2},0,0,0,0,0,0,0} , 
\func{}{\tfrac{k-1}{2},0,0,0,0,0,1,0} , 
\func{}{\tfrac{k-3}{2},0,1,0,0,0,0,0} , \\
\func{}{\tfrac{k-5}{2},0,0,0,1,0,0,0} , 
\func{}{\tfrac{k-5}{2},0,1,0,0,0,1,0} , 
\func{}{\tfrac{k-5}{2},0,0,0,0,2,0,0} , \\
\func{}{\tfrac{k-7}{2},0,0,0,0,2,0,0} , 
\func{}{\tfrac{k-7}{2},0,1,0,0,0,2,0} , 
\func{}{\tfrac{k-9}{2},0,0,0,0,1,0,2} , \\
\func{}{\tfrac{k-9}{2},0,1,0,1,0,0,0} , 
\func{}{\tfrac{k-9}{2},1,0,0,0,1,0,1} , 
\func{}{\tfrac{k-11}{2},0,0,0,0,0,6,0} , \\
\func{}{\tfrac{k-11}{2},0,0,0,2,0,0,0} , 
\func{}{\tfrac{k-13}{2},0,0,0,1,0,4,0} , 
\func{}{\tfrac{k-15}{2},0,0,0,0,4,0,0} , \\
\func{}{\tfrac{k-15}{2},0,0,2,0,0,3,0} , 
\func{}{\tfrac{k-17}{2},0,0,0,0,0,0,6} , 
\func{}{\tfrac{k-17}{2},0,0,0,3,0,0,0} , \\
\func{}{\tfrac{k-17}{2},0,1,0,1,2,0,0} , 
\func{}{\tfrac{k-19}{2},0,1,2,0,0,0,2} , 
\func{}{\tfrac{k-19}{2},0,2,0,2,0,0,0} \} \,.
\end{multline*}
For $k$ even,
\begin{multline*}
\{ \func{}{\tfrac{k-2}{2},0,0,0,0,0,0,1} , 
\func{}{\tfrac{k-2}{2},1,0,0,0,0,0,0} , 
\func{}{\tfrac{k-4}{2},0,0,1,0,0,0,0} , \\
\func{}{\tfrac{k-6}{2},0,0,0,0,1,0,1} , 
\func{}{\tfrac{k-6}{2},1,0,0,0,0,2,0} , 
\func{}{\tfrac{k-8}{2},0,0,0,0,0,0,3} , \\
\func{}{\tfrac{k-8}{2},0,0,0,1,0,0,1} , 
\func{}{\tfrac{k-8}{2},0,0,1,0,0,2,0} , 
\func{}{\tfrac{k-8}{2},3,0,0,0,0,0,0} , \\
\func{}{\tfrac{k-10}{2},0,1,0,0,1,0,1} , 
\func{}{\tfrac{k-10}{2},0,0,0,1,0,1,1} , 
\func{}{\tfrac{k-10}{2},1,0,1,0,0,0,1} , \\
\func{}{\tfrac{k-12}{2},0,0,0,0,0,5,1} , 
\func{}{\tfrac{k-12}{2},0,0,1,1,0,1,0} , 
\func{}{\tfrac{k-14}{2},0,0,1,0,1,3,0} , \\
\func{}{\tfrac{k-16}{2},0,0,1,0,3,0,0} , 
\func{}{\tfrac{k-16}{2},0,1,1,0,1,2,0} , 
\func{}{\tfrac{k-18}{2},0,0,2,0,0,0,3} , \\
\func{}{\tfrac{k-18}{2},0,1,0,0,0,0,5} , 
\func{}{\tfrac{k-18}{2},0,1,0,2,0,0,1} , 
\func{}{\tfrac{k-18}{2},0,2,0,0,2,0,1} \} \,.
\end{multline*}
\item[$\boldsymbol{F_4}$]  For $k$ odd,
\begin{multline*}
\{ \func{}{\tfrac{k+1}{2},0,0,0} , \func{}{\tfrac{k-1}{2},0,0,2} , 
\func{}{\tfrac{k-1}{2},0,1,0} , \func{}{\tfrac{k-3}{2},1,0,1} , \\
 \func{}{\tfrac{k-3}{2},0,2,0} , \func{}{\tfrac{k-5}{2},0,1,4} , 
\func{}{\tfrac{k-5}{2},1,0,3} \} \,.
\end{multline*}
For $k$ even,
\begin{multline*}
\{ \func{}{\tfrac{k}{2},0,0,1} , \func{}{\tfrac{k-2}{2},0,1,1} , 
\func{}{\tfrac{k-2}{2},1,0,0} , \func{}{\tfrac{k-4}{2},0,0,5} , \\
\func{}{\tfrac{k-4}{2},1,0,2} , \func{}{\tfrac{k-4}{2},0,2,1} , 
\func{}{\tfrac{k-4}{2},1,1,0} , \func{}{\tfrac{k-6}{2},1,0,4} \} \,.
\end{multline*}
\item[$\boldsymbol{G_2}$]  For $k$ odd,
$$
\{ \func{}{\tfrac{k+1}{2},0} , \func{}{\tfrac{k-1}{2},2} , 
\func{}{\tfrac{k-3}{2},4} \} \,.
$$
For $k$ even,
$$
\left\{ \func{}{\tfrac{k}{2},1} , \func{}{\tfrac{k-2}{2},3} \,, 
\func{}{\tfrac{k-4}{2},5} \right\} \,.
$$
\end{description}

\section{Proof using fusion potentials} \label{appfuspot}

It is well-known that in the case of fusion rings, the 
affine algebraic variety defined
by the fusion ideal is a so-called Jacobian variety, i.e., if
$\cF = \CC[x_1,\ldots,x_N]/\cI$ we can find a so-called fusion
potential $V(x_1,\ldots,x_N)$ such that 
$\cI = \langle P_1,\ldots,P_N\rangle$,
where $P_i \equiv \partial V / \partial x_i$ 
\cite{Ge,MRS}--\cite{GS,Cre}.
Since the fusion potential provides a set of generators of the fusion
ideal, it can be used to give an alternative proof of \eqref{eqCaa}.
While an expression 
for the fusion potential is known, in principle, for all compact,
simple, connected, simply-connected Lie groups $G$ \cite{Cre}, 
we have not been 
able to make effective use of it for cases other than $A_n$ and $C_n$. 
In this appendix we present an alternative proof for those two cases.

The fusion potential for $A_n$ takes the form \cite{Ge}
\begin{equation} \label{eqAppBa}
  \func{V}{\chi_1, \ldots, \chi_n} = \frac{1}{k+n+1} \sum_{i=1}^{n+1}
  q_i^{k+n+1}\,,
\end{equation}
where the (formal) variables $q_i$ are related to the standard 
(overcomplete) basis $\{\epsilon_i\}_{i=1}^{n+1}$ for the weight space
of $A_n$ by $q_i = \exp(\epsilon_i)$, with the constraint 
$q_1 \cdots q_{n+1} = 1$.
In terms of these variables the 
characters $\chi_m$ of the fundamental representations $L(\Lambda_m)$,
($m=1,\ldots,n$), are given in terms of the elementary symmetric
polynomials, i.e.\
\begin{equation} \label{eqAppBb}
  \chi_m = \sum_{1 \leqslant i_1 < \ldots < i_m
  \leqslant n+1} q_{i_1} \cdots q_{i_m} \,.
\end{equation}
Defining
\begin{equation} \label{eqAppBc}
  V_m = \frac{1}{m} \sum_{i=1}^{n+1} q_i^m \,,
\end{equation}
we can form the generating function
\begin{equation} \label{eqAppBd}
  \func{V}{t} = \sum_{m=1}^{\infty} \func{}{-1}^{m-1} V_m\, t^m = 
  \log \left( \sum_{i=0}^{n+1} \chi_i\, t^i \right) \,,
\end{equation} 
where we have set $\chi_0=1=\chi_n$ for convenience.  The fusion potential
is now (up to a sign) the coefficient of $t^{k+n+1}$ in $V(t)$, and the
generators of the fusion ideal are the derivatives of this coefficient
with respect to the $\chi_i$ ($i = 1, \ldots, n$).  Differentiating
$\func{V}{t}$ with respect to $\chi_j$, and evaluating at $q_i=1$
(i.e.\ at the point where $\chi_i = d_{\Lambda_i} = \binom{n+1}{i}$), 
gives
\begin{equation} \label{eqAppBe}
  \left. \pardiff{\func{V}{t}}{\chi_j} \right|_{\chi_i = d_{
  \Lambda_i}} =   \frac{t^j}{(1+t)^{n+1}} \,.
\end{equation} 
Taylor expanding this around $0$ and extracting the coefficient of
$t^{k+n+1}$ gives the values of the generators of the fusion ideal 
under the dimension map, and hence determines $x$. The result is
\begin{equation} \label{eqAppBf}
  x = \gcd \set{\binom{k+2n+1-j}{n} : j = 1,\ldots,n} = 
  \gcd \set{\binom{k+n+j}{n} : j = 1,\ldots,n}  \,,
\end{equation} 
in accordance with Eqn.~\eqref{eqDAab}.

The proof above for $A_n$ easily generalizes to $C_n$.  The fusion
potential is given by \cite{BMRS,GS}
\begin{equation} \label{eqAppBg}
  \func{V}{\chi_1,\chi_2,\dots,\chi_n} = \frac{1}{k+n+1} 
  \sum_{i=1}^n \func{}{q_i^{k+n+1} + q_i^{\func{-}{k+n+1}}} \,,
\end{equation} 
where, in terms of an orthonormal basis $\{ \epsilon_i\}_{i=1}^n$
for the weight space of $C_n$, the variables $q_i$ are given
by $q_i = \exp(\epsilon_i)$, and 
the characters $\chi_j$ ($j=1,\ldots,n$)
are related to the variables $q_i$ by
\begin{equation}  \label{eqAppBh}
  \chi_j  = E_j - E_{j-2} \qquad \text{($E_j = 0$ when $j<0$),} 
\end{equation}
and $E_j$ is defined by
\begin{equation}  \label{eqAppBi}
  \sum_{j=0}^{\infty} E_j t^j  = \prod_{i=1}^n 
  \func{}{1 + q_i t} \func{}{1 + q_i^{-1} t}\,.
\end{equation}
In particular we have $E_j = E_{2n-j}$, and $E_j = 0$ for $j>2n$.
Another consequence
of this is that $\chi_j + \chi_{2n+2-j} = 0$, so there are indeed only
$n$ linearly independent characters (in agreement with the arguments
of the fusion potential $V$).

In analogy to $A_n$, we define $V_m = \frac{1}{m} \sum_{i=1}^n
\func{}{q_i^m + q_i^{-m}}$, and form the generating function
\begin{equation} \label{eqAppBj}
  \func{V}{t} = \sum_{m=1}^{\infty} \func{}{-1}^{m-1} V_m \, t^m = 
  \log \left(\sum_{j=0}^{2n} E_j \, t^j \right)\,. 
\end{equation}
Differentiating with respect to the $\chi_i$ gives
\begin{equation}\label{eqAppBk}
  \pardiff{\func{V}{t}}{\chi_i} = \frac{t^i + t^{i+2} + \ldots +
  t^{2n-2-i} + t^{2n-i}}{\sum_{j=0}^{2n} E_j t^j} \,.
\end{equation}
Evaluating this at $q_i=1$, i.e.\ at the point 
$\chi_i = d_{\Lambda_i} = \binom{2n}{i} - \binom{2n}{i-2}$
(thus $E_i = \binom{2n}{i}$), we find
\begin{equation}\label{eqAppBl}
  \left. \pardiff{\func{V}{t}}{\chi_i} \right|_{\chi_j = d_{\Lambda_j}} = 
  \frac{t^i + t^{i+2} + \ldots + t^{2n-2-i} + t^{2n-i}}
  {\func{}{1 + t}^{2n}} \,.
\end{equation}
Taylor expanding about $0$, and extracting the coefficient of
$t^{k+n+1}$, gives the charge group parameter as
\begin{align*}
\begin{split}
x & = \gcd \set{\sum_{j=0}^{n-i} \binom{k+3n-i-2j}{2n-1} : i = 1 , 
\ldots , n} \\
& = \gcd \set{\sum_{j=0}^{i-1} \binom{k+2n+1-i+2j}{2n-1} : i = 1 , 
\ldots , n} \\
& = \gcd \left\{ \binom{k+2n}{2n-1} , \binom{k+2n+1}{2n-1} + 
\binom{k+2n-1}{2n-1} , \right. \\
& \qquad \left. \binom{k+2n+2}{2n-1} + \binom{k+2n-2}{2n-1}, 
\ldots , \binom{k+3n-1}{2n-1} + \binom{k+n+1}{2n-1} \right\} \,,
\end{split}
\end{align*}
as in Eqn.~\eqref{eqDBbb}.


\end{document}